\documentclass[11pt]{article}
 \usepackage{mor1,epsfig}

\newcommand{\SM}{SU(3) \otimes SU(2) \otimes U(1)}

\def\be{\begin{equation}}
\def\ee{\end{equation}}
\def\bea{\begin{eqnarray}}
\def\eea{\end{eqnarray}}
\def\bi{\begin{itemize}}
\def\ei{\end{itemize}}

\begin{document}
\vspace*{2cm}
\title{ANOMALY CONSTRAINTS AND NUMBER OF FAMILIES}

\author{ Y. GOUVERNEUR }

\address{Universit{\'e} Libre de Bruxelles, D\'epartement de Physique Th\'eorique CP 225,  2 Blvd du Triomphe,\\
Brussels 1050, Belgium}

%\vspace*{4cm}

\maketitle
\vspace*{2cm}
\abstracts{The purpose of this paper is to answer the following question : is it possible to find a gauge group G and a space-time dimension D such that anomalies constrain the fermion content to be composed of the 3 known families of quarks and leptons ? The answer is, in some way, unique and consists in taking an $SU(5)$ gauge group in 6 dimensions. This work has been done with the collaboration of N. Borghini and M. H. Tytgat.}

Reading the above question, one could or even should immediately worry whether it's really worth involving Grand Unified Theories (GUT) and extra dimensions. The point is that in the standard model, fermions have a very particular quantum numbers pattern, coming in three copies called families or generations. And there is no reason for this. The point of view of this paper is to show that this result can be the only possibility left if we require consistency (through anomaly cancellations) of a six dimensional GUT theory.
The paper is organized as follows.

The first two sections are devoted to a review on the family structure in the standard model and to the different types of anomalies. The next section studies the magic of anomalies in the usual $\SM$ standard model in 4 dimensions. This magics will be a strong motivation for looking beyond the standard model. Finally, we'll make some comments on extensions and phenomenology.

Many technical details will be skipped here, especially the Green-Schwarz mechanism won't be described. On the other hand, many ``well-known'' points will be stressed on, in order to emphasize that their validity is restricted to 4 dimensions. A more complete version of the calculations and discussions can be found in \cite{bgt}.  

\section{Review : Standard Model fermion content}
Particles are described by fields, because it's the only consistent with quantum mechanics and special relativity way to do it \cite{weinbergQFT}. These fields can be written in terms of creation an annihilation operators which correspond to particles and/or to antiparticles. We'll restrict our attention to the fermionic sector of the standard model, composed of quarks and leptons (and their corresponding antiparticles), with yet unbroken $\SM$ gauge theory. So fermions are massless. In fact, spontaneous symmetry breaking (SSB) to $SU(3) \otimes U(1)$ will not affect our results, because anomalies are insensitive to SSB. 

The fermionic particles and antiparticles can all be described by $3 \times 15$ fermionic fields. These particles have specific gauge interactions and can be classified according to the triplet, anti-triplet or singlet (3,$\bar{3}$ or 1), the doublet or singlet (2 or 1) and their hypercharge ($Y$) representations of $SU(3)$, $SU(2)$ and $U(1)$ respectively. 
\vspace{0.5cm}
\begin{center}
\begin{tabular}{|c|c|c|c|c|c|c|} \hline
& \multicolumn{3}{c|}{3 fermion generations} & \multicolumn{3}{c|}{$\SM$} \\ \hline
\rule[-8mm]{0mm}{15mm} $\mathcal{Q}$ &  $ \left( \begin{array}{c}
\rm{u} \\ \rm{d}
\end{array}\right)_{\! L} $ &  $ \left( \begin{array}{c}
\rm{c} \\ \rm{s}
\end{array}\right)_{\! L} $ &  $ \left( \begin{array}{c}
\rm{t} \\ \rm{b}
\end{array}\right)_{\! L} $  & \hspace{7mm} 3 \hspace{7mm} & \hspace{7mm} 2 \hspace{7mm} & 1/3 \\ \hline
\rule[-5mm]{0mm}{10mm} $\mathcal{U}$ & ${(\rm{u}^{c})}_{\, L}$ & $ {(\rm{c}^{c})}_{\, L}$ & $ {(\rm{t}^{c})}_{\, L}$ & $\bar{3}$ & 1 & -4/3 \\ \hline
\rule[-5mm]{0mm}{10mm} $\mathcal{D}$ & ${(\rm{d}^{c})}_{\, L}$ & $ {(\rm{s}^{c})}_{\, L}$ & $ {(\rm{b}^{c})}_{\, L}$ & $\bar{3}$ & 1 & 2/3 \\ \hline
\rule[-8mm]{0mm}{15mm} $\mathcal{L}$ &  $ \left( \begin{array}{c}
\nu_{e} \\ e
\end{array}\right)_{\! L} $ &  $ \left( \begin{array}{c}
\nu_{\mu} \\ \mu
\end{array}\right)_{\! L} $ &  $ \left( \begin{array}{c}
\nu_{\tau} \\ \tau
\end{array}\right)_{\! L} $  & 1 & 2 & -1 \\ \hline
\rule[-5mm]{0mm}{10mm} $\mathcal{E}$ & ${(\rm{u}^{c})}_{\, L}$ & $ {(\rm{c}^{c})}_{\, L}$ & $ {(\rm{t}^{c})}_{\, L}$ & 1 & 1 & 2 \\ \hline
\end{tabular}
\end{center}
\vspace{0.5cm}

A crucial point is that all (anti-)leptons and (anti-)quarks can be described with these left-handed chiral fields. A Dirac spinor is composed of a left and a right Weyl spinor field. This chirality is related to the eigenvalues of the $\gamma_5$ matrix which commutes with all Lorentz transformations (and so these eigenvalues are good quantum numbers). Chirality of the fields is directly linked to helicity (projection of spin in direction of propagation) of the particle the field describes. This is also true in all other even dimensions, where the chirality notion can be extended. Dirac spinors can be defined in all $D$ space-time dimensions, and have increasing number of components ($2^{D/2}$ if $D$ is even).

Charge conjugation flips chirality (in four dimensions), and so one could have chosen to take (for example) right-handed $u_{R}$ field instead of $u^{c}_{L}$ field : they both describe a ``right'' helicity $u$-quark particle and a ''left'' helicity $u$-antiquark. The two descriptions will not necessarily be equivalent anymore in other space-time dimensions.

This rather complicated family structure has motivated people to try incorporating it in bigger representations of GUT groups containing the standard model $\SM$. The minimal one is $SU(5)$ and fermions of each family can be nicely classified by left-handed $\bar{5}$ and $10$ representations :

\vspace{0.2cm}
\begin{center}
\begin{tabular}{|c|c|} \hline
$SU(5)$ & $\SM$ \\ \hline
\rule[-5mm]{0mm}{10mm} $\bar{5}$ & $\mathcal{D} + \mathcal{L}$ \\ \hline
\rule[-5mm]{0mm}{10mm} $10$ & $\mathcal{Q} + \mathcal{U} + \mathcal{E}$ \\ 
\hline
\end{tabular}
\end{center}

\section{Review : Anomaly species and number of dimensions}
A symmetry is said to be anomalous if it exists at the classical level, 
but does not survive quantization. 
In some cases, anomalies are welcome, as in the study of the $\pi^0$ decay.

Other anomalies, in opposition, affect local symmetries, in particular gauge 
symmetries, and jeopardize the theory consistency.
Such anomalies spoil renormalizability; but even in the case of effective, 
{\it a priori} non-renormalizable theories, they destroy unitarity, leading to 
theories without predictive power. 
Consistent models should therefore either contain none of these anomalies, 
or automatically cancel them.
Conversely, the cancellation condition gives useful constraints on the theory's matter content.

We'll restrict on these chiral gauge (or gravitional) anomalies, which occur when chiral fermions couple to gauge fields and/or gravitons. Their presence is subject to the very existence of chirality, which can only be defined properly in even space-time dimensions.

\subsection{Local anomalies}
They arise from a typical kind of Feynman diagram with a specific number ($D/2+1$) of 
internal arcs (made of chiral fermions), depending on the spacetime dimension $D$. 
In 4 dimensions, these are the well-known triangle diagrams. 
In 6-, 8-, and 10-dimensional theories, the corresponding possibly anomalous 
diagrams are respectively the so-called box, pentagonal, and hexagonal 
diagrams. Requiring the absence of anomalies translates thus in
 the cancellation (considering all relevant fermions) of these diagrams. 

We shall distinguish three types of local anomalies, according to the nature 
of the external legs of the anomalous diagrams. 
Diagrams with only gauge bosons (resp. gravitons) correspond to the pure gauge (resp. gravitational) anomaly \cite{Alvarez-Gaume:1984ig}. 
Finally, the mixed anomaly correspond to diagrams with both gauge bosons and 
gravitons. \cite{Alvarez-Gaume:1984ig}

In the pure gauge case, the anomaly is proportional to a group factor, 
which multiplies a Feynman integral: 
\begin{equation}
\label{gaugeDD}
\sum_{L_D} {\rm STr} \left( T^a T^b \ldots T^{\frac{D}{2}+1} \right) - 
\sum_{R_D} {\rm STr} \left( T^a T^b \ldots T^{\frac{D}{2}+1} \right),
\end{equation}
where the notation ${\rm STr}$ means that the trace is performed over the 
symmetrized product of the gauge group generators $T^a$. The latter depend on the representation of the gauge the fermions belong to.
All representations can be constructed in terms of fundamental representations (noted $t^a$ below). So traces need not to be calculated for all cases. Moreover, traces can be decomposed in terms of irreducible ones. The sums run over all left- and right-handed (in the $D$-dimensional sense) fermions of the theory belonging to the representation $T^a$ . 

A necessary and sufficient condition for the absence of local gravitational 
anomaly is therefore the identity of the numbers of left- and right-handed 
fermions:
\begin{equation}
\label{gravitational}
N_{L_D}-N_{R_D}=0. 
\end{equation}

There is a subtlety here. In dimension $D=4k$ charge conjugation 
${\sf C}$ flips chirality, while it does not in $D=4k+2$. 
Therefore, a left-handed Weyl fermion field describes a left-handed (helicity) particle 
and an antiparticle with opposite (resp.\ identical) chirality (helicity) in $D=4k$ 
(resp.\ $D=4k+2$).
Such a field is, from the gravitational point of view, vector-like in dimension 
$4k$, while it is chiral in dimension $4k+2$. 
Thus, the local gravitational anomaly automatically vanishes if the spacetime 
dimension is $D=4k$. 

The mixed gauge-gravitational anomaly is proportional to the product of a gauge group factor 
(similar to the pure gauge case, but with an appropriate number of generators) and a 
gravitational term \cite{Alvarez-Gaume:1984ig}.
This latter vanishes when the number of gravitons is odd. 

When the mixed anomaly does not vanish thanks to group properties or an 
appropriate fermion choice, it may still be canceled through the Green-Schwarz 
mechanism.

\subsection{Witten's Global anomalies}

The global gauge anomaly \cite{Witten:1982fp} occurs in presence of chiral fermions, when gauge 
transformations which cannot be deduced continuously from the identity exist.
The anomaly then leads to mathematically inconsistent theories in which all 
physical observables are ill-defined. It vanishes only if the matter content of the theory is appropriate. 
More precisely, if the $D$-th homotopy group of the gauge 
group $G$, $\Pi_{D}(G)$, is nontrivial i.e. $\Pi_{D}(G)\neq 0$, the cancellation of the anomaly 
constrains the numbers $N(p_{L_D})$ and $N(p_{R_D})$ of left- and right-handed 
$p$-uplets is : 
\begin{equation}
\label{global}
\Pi_D(G)=Z_{n_D} \Rightarrow  
c^p_D \left[N(p_{L_D})-N(p_{R_D})\right]=0 \mbox{ mod } n_D,
\end{equation}
where $c^p_D$ is an integer whose value depends on the spacetime dimension $D$, 
the gauge group $G$, and the representation $p$ of $G$ the fermions belong to.

\section{Standard Model in 4 dimensions vs Anomalies}

As an useful exercise, one can check that with the fermion content of the standard model, all anomalies
 (in 4 dimensions) cancel automatically.

Let us begin with the perturbative anomalies, related to triangle diagrams see ~Fig.~1.

\begin{figure}
\begin{picture}(0,100)%
\centerline{\epsfxsize=0.3\textwidth \epsfbox{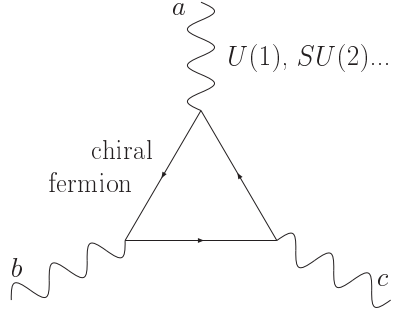}}
\end{picture}%
\caption{ Each vertex describes the interaction of a chiral fermion in representation $T^a$ and a gauge field
\label{fig:at2}}
\end{figure}

Absence of pure local gauge anomaly brings several constraints, due to the presence of three gauge groups in $\SM$
 : each of the three vertices can carry a gauge field of one of the three groups. One applies eq.(\ref{gaugeDD}) for each case :  
 \bi 
  \item $SU(3)^3$ : $ \sum_{L} {\rm STr}\, (T^a T^b T^c) = (2-1-1) \times {\rm STr}\, (t^a t^b t^c) $
  \item $SU(2)^3$ : $ {\rm Tr}\,(\sigma^a \{\sigma^b , \sigma^c) \} = 2\delta^{bc}\,{\rm Tr}\,\sigma^a = 0$
  \item $U(1)^3$ : $\sum_{L} Y^3 = 6 \left(\frac{1}{3} \right)^3 + 3 \left(\frac{-4}{3}\right)^3 + 3 \left(\frac{2}{3} \right)^3 + 2 (-1)^3+ 2^3 = 0$
  \item $SU(3)^2\,U(1)$ : $ \sum_{\bf 3,{\bf \bar 3}} Y = 2\times \frac{1}{3} + \frac{-4}{3} + \frac{2}{3} = 0 $
  \item $SU(2)^2\,U(1)$ : $ \sum_{\rm doublets} Y = 3\times \frac{1}{3} + (-1) = 0 $
  \ei

The $\sigma$ matrices are the usual Pauli matrices.
The notation $SU(3)^3$ simply means that one has a gluon at each vertex. Let look a bit more $SU(3)^3$ : there is a doublet of left-handed triplets (factor 2) and two singlets of anti-triplets (two -1 factors). These -1 factors come from the fact that generators are hermitian, and the trace of three conjugate $T^*$ is equal to -1 the trace over three $T$. The first line cancellation just reflects the fact that $SU(3)$ is vector-like in the standard-model.
 
It is a remarkable feature that the third cancellation condition requires presence of both leptons and quarks. It would be nice is something similar happens between fermions in different generations.

The mixed anomaly is related to one graph with one gauge field and two gravitons. The cancellation condition reads :
\be
 \sum_{ L} \rm{Tr} Y = 6\times \frac{1}{3} + 3 \times \left(\frac{-4}{3}\right) + 
3\times \frac{2}{3} + 2 \times (-1) + 1 \times 2 = 0
\ee

If the hypercharge was some generator of a bigger semi-simple group, than this condition would be trivial (generators are traceless). 
This can be seen a a motivation for GUTs.

The only group out of $\SM$ which has non-trivial homotopy (in 4 dimensions) and thus potentially dangerous for global gauge anomaly is $SU(2)$. To avoid it, the number of doublets (left minus right) should be even, and it is indeed :
\be N(2_L)=3+1=4=0 \mbox{ mod } 2
\ee

As said before (we are in 4 dimensions) there are no gravitational anomalies. So all anomalies cancel, some in a quite miraculous way, between fermions of each generation. So anomalies don't give any informations on the number of generations. Looking in other space-time dimensions would bring other anomaly constraints which maybe can say something about genrations. By the way, the magic of these cancellation can be understood (in 4 dimensions) if one remarks that all fermions of a generation fit in a $\bar{5}+10$ representation of SU(5) or a 16 of SO(10)...

\section{SU(5) in 6 dimensions vs Anomalies}

In 6 dimensions, many features change : there are gravitational anomalies and self-dual tensors exist and do contribute to anomalies. The latter mechanism is the so-called Green-Schwarz mechanism. This mechanism unables to cancel specific mixed anomalies and some parts (called reducible) of pure gauge anomalies.

We'll consider our fermions to belong to the representations $\bar{5}$ and 10 of $SU(5)$ in order to be able to reproduce at the end the standard model content.

The 10 representation is in fact constructed from the anti-symmetric part of the product of two 5 generations. So traces over its generators can be expressed in terms of traces over generators $t^a$ of the fundamental 5 representation. The pure gauge anomaly group coefficient for the relevant box diagram writes now simply : 
\bi
\item{$ {\rm STr}\, (T^4) \;  = 1 \times {\rm STr}\, (t^4)$}
\item{$  {\rm STr}\, (T^4) \;  = -3 \times {\rm STr}\, (t^4) + 3\times {\rm STr}\, (t^2)^2$}
\ei
The factorized part of the trace ${\rm STr}\, (t^2)^2$ is reducible and can be cancelled by introducing an appropriate self-dual tensor.
So, the most economical chiral solution  consists of  ${\bf\bar 5}_L$, 
${\bf\bar 5}_L$, ${\bf\bar 5}_L$, ${\bf 10}_L$. Of course, as chiral anomalies can only constrain chiral fermions, any vector-like solution ${\bf\bar 5}_L$, ${\bf 10}_L$, ${\bf\bar 5}_R$, ${\bf 10}_R$... is allowed.
But in the standard model one has 45 fermions in generations, so one should add quantum numbers if one hopes to complete our minimal solution to describe all fermions in the standard model. Let us thus add $ {\bf 10}_L, {\bf 10}_R $. So the standard model content seems to be the minimal chiral anomaly free solution in 6 dimension, plus a vector-like component.
So anomalies cannot tell us why there are generations. But if you want some copies of the quantum numbers corresponding to the ones you see per family in nature, then they should come 3 by 3.

As said before, the mixed anomaly can be killed by Green-Schwarz mechanism. The relevant self-dual tensor contributes to the gravitational anomaly, and forces us to introduce 3 gauge chiral singlets.

As $\Pi_{6}(SU(5))=0$, there is no global gauge anomaly.

\section{Extensions - Phenomenology}

Possible extensions are numerous. Some obvious ones consist in taking other GUT groups as $SO(10)$ and $E_6$. This has been done in our paper \cite{bgt}. But it is only $SU(5)$ which brings the number 3 of generations. A nice feature of our minimal solution is that it can, in some way, be put in a single 27 representation of $E_6$, at the price of introducing other fermions which have not been observed. One can also try to supersymmetrize the model, the price is once again the appearance of many other fields, loosing in simplicity. Nevertheless, it would bring a more unified approach, because our solution with fermions has brought (without supersymmetry) bosonic fields (the Green-Schwarz tensor).

To study the phenomenology of our solution, one has to say something about the extra dimensions and how one recovers 4 dimensional predictions. One crucial problem (which is a generic problem) is that a chiral fermion in 6 dimensions is composed of a left and a right-handed 4-dimensional fermion. So some chirality selection mechanism (coupling to a vortex, orbifold boundary conditions ...) has to be implemented, and will affect strongly and in a model-dependant way the phenomenology. 
One could nevertheless make some general remarks on allowed mass and Yukawa couplings in $D=6$. There are two constraints : six-dimensional Lorentz invariance and $SU(5)$ Higgs mechanism. Our solution is :

\[
\left( \begin{array}{c}
{\bf\bar 5}_L \\ {\bf 10}_L
\end{array}\right), \quad 
\left( \begin{array}{c}
{\bf\bar 5}_L \\ {\bf 10}_L
\end{array}\right), \quad 
\left( \begin{array}{c}
{\bf\bar 5}_L \\ {\bf 10}_R
\end{array}\right)
\]

Lorentz invariance requires a Yukawa coupling term to contain a left field and a right field (in 4 dimensions, chirality flips charge conjugation, and one can write such a term even if one has only left fields). $SU(5)$ invariance and Higgs mechanism further restricts possible terms to be of two kinds : ${\bf 10}_L$ + ${\bf 10}_R$ which could give rise to the top mass, and ${\bf\bar 5}_L$ + ${\bf 10}_R$ giving the bottom mass. Kaluza-Klein corrections should be smaller than this values. This would be a way to understand why one generation is much heavier than the others.

Some other interesting features could come from quantum corrections and dimensional reduction. One can study CP violation because Yukawa couplings are complex, and also find some solution to the strong CP problem thanks to the Green-Schwarz tensor \cite{Fabbrichesi:2001fx}.

Even the problem of proton stability could be adressed and solved by using a residual (in 4 dimensions) discrete Lorentz invariance, see \cite{Appelquist:2001mj}.

\section*{Acknowledgements}

Y. G. benefits from an F.N.R.S. grant.  Special thanks to the organizers of the 37th Moriond session ``Electroweak interactions and unified theories'' for invitation, possibility to attend, and kind hospitality during the conference. Y.G. would like to thank G.~Tasinato for insightful discussions on Green-Schwarz mechanism. This work was done in collaboration with N. Borghini and M. H. Tytgat.

\end{document}